\documentclass[a4paper,twocolumn]{article}

\usepackage{graphicx}

\title{Reducing Bias and Optimising Execution Time in Iterative Solutions of the Time Dependent Ginzburg Landau Equations}
\author{E. R. Di Lascio \\ \scriptsize{erlascio@protonmail.com}}
\date{\today}

\begin{document}
\twocolumn[
\begin{@twocolumnfalse}
\maketitle
\abstract
The importance of simulating pinning arrays in superconducting samples for the increase of critical currents has been increasing in the last few years. Since the Time Dependent Ginzburg Landau (TDGL) can be more accurate than alternative methodologies, the simulation procedures involving it are critical to design devices that can sustain higher critical currents and, therefore, to the field of applied superconductivity. In this article, a simple novel algorithm is presented for the reduction of bias and optimisation of execution time in iterative time dependent simulations, applied to TDGL solutions of superconducting samples. Taking a time series approach to the magnetic response of the sample, stationary solutions are found for each step in the evolution of the applied external field, leading to bias reduction and minimisation of iterations needed to be spent at each step in the applied field. The results are presented for a pure superconductor, in a framework of simulations via link variable technique, with simple Euler algorithm for the solution in time, but the implementation can be adapted easily to deal with adaptive step size solutions or semi-implicit methods, which are not exempt from the bias and iterations tradeoff. 
\\ \ \\ \ \\ \ 
\end{@twocolumnfalse}]

\section{Introduction}
\label{intro}

The field of applied superconductivity is continuously benefited from improvements in sample preparations that lead to greater critical currents. As examples, new thin film samples standing higher critical currents have been developed, and the extension to superconducting magnets could follow.

The importance of simulations to the goal of optimisation of critical currents is ever increasing, as the pinning arrays in superconducting samples show promising results and the increasing complexity of the pinning centres arrangements demand multiplied efforts. This task is far more easily accomplished computationally. 

In order to meet the challenge of providing accurate results to optimisation of critical currents, the methodology has to be chosen carefully and its limitations have to be taken into account.

The effectiveness of the application of the Time Dependent Ginzburg Landau (TDGL) equations to the description of superconductors and hybrid samples is well established for a long time now and still they preserve a great relevance and are vastly employed in current research, with examples including comparisons of numerical with experimental results \cite{lascio}, \cite{rede}.

The TDGL equations are coupled partial differential equations for both the magnetic potential and the superconducting order parameter. They have many numerical methods of solution. This article focuses on one method which has been widely applied, the link variable technique (see, for example, a short review in \cite{lascio}, the original in \cite{kato} with another version in \cite{gropp}), but the results presented here should be applicable to other iterative methodologies.

It is a quite efficient and flexible methodology, allowing for various parameterisations. It is however relatively high in time consumption when applied to calculation of magnetic response of a sample to a time dependent applied field, partially due to the many iterations required for the physical accuracy and numerical stabilisation at each step of the evolution of field or current. So it is desirable (and indeed feasible) to reduce the number of iterations spent on stabilisation.

On the other hand, the lack of sufficient iterations until stabilisation at each step in the applied field leads to biases in the estimation of magnetic response, leading to unrealistic conclusions.

The iterative numerical simulations of time dependent physical systems depend, among many other things, on the time interval for each iteration, as well as on the time scales of the internal system and of the external influence.  

The efficiency of a numerical algorithm to solve the differential equations lead to less time comsumption indeed, but nonetheless, to achieve numerical stabilisation of the solution throughout the sample and indeed to accurately infer the magnetic response of the sample, many iterations (however efficient) will be spent. This is so because each iteration represents a discrete step in time. Being so, and knowing that the physical nature of the sample (subject to magnetic after effect, for example) requires a finite time interval to present a stable magnetic response, as well as the finite size of the sample requires that a finite amount of time be spent on the propagation of a new value of applied magnetic field from the border to the centre of the sample, no matter how efficient a numerical algorithm for the solution of the differential equations is, it is necessary to spend iterations in the physical and numerical stabilisation of the system, at each new value of the applied field. 

The original algorithm \cite{kato} proposed the use of $10^5$ iterations at each step in the magnetic field evolution as a sufficient quantity for the numerical stabilisation.

In order to improve accuracy in simulated results, this article studies the extent to which the choice of step can lead to bias in the results, and an alternative choice criterion is proposed. The main goal is to provide accurate results in the description of superconducting and hybrid samples, enabling a realistic optimisation of critical currents, thus concretely contributing to applied superconductivity. 

This article presents results of a study on the necessary number of iterations until stabilisation in a variety of cases, allowing the definition of some requirements on criteria for finishing any step and proceeding to the next in the field or current evolution. The consequences of applying insufficient number of iterations are explored.

After obtaining these results, some possible criteria are discussed, leading to the choice of one which is simple and effective, resulting in a novel algorithm for identification of stabilisation. This is then applied to the study of the magnetisation of the sample and the critical current evolution as function of the magnetic field. 

The results are compared with some other results originated from the application of the simple original criterion, demonstrating the significance of the proposed algorithm for both the efficiency of the execution and for the reliability of the results, avoiding biases in the curves of magnetisation and critical current and simultaneously optimising the number of iterations.  

\section{Specification}
\label{sec2}

The TDGL equations are coupled partial differential equations for both the magnetic potential and the superconducting order parameter. Here, only the solutions will be presented, as they are necessary for implementation of the algorithm and study of the physical properties. The relevant steps to obtaining these solutions can be found in reference \cite{lascio}. 

The dependence in time is expanded iteratively. Using the simple Euler method for the solution in time, the iterative solutions are given by:

\begin{equation}
\Delta _{j} (t + \delta t) = \Delta _{j}(t) + \mathcal{F} _{\Delta} ^{j} (t) \delta t,
\label{delta_it_eq}
\end{equation}
\begin{equation}
U _{x} ^{jk} (t + \delta t) = U _{x} ^{jk} (t) \exp(\mathcal{F} _{U _{x}} ^{jk} (t) \delta t),
\label{ux_it_eq}
\end{equation}
\begin{equation}
U _{y} ^{jm} (t + \delta t) = U _{y} ^{jm} (t) \exp(\mathcal{F} _{U _{y}} ^{jm} (t) \delta t),
\label{uy_it_eq}
\end{equation}
where:
\begin{eqnarray}
\mathcal{F} _{\Delta} ^{j} (t) \equiv \frac{1}{12}\Bigg[\frac{U _{x} ^{kj} (t) \Delta _{k} (t) - 2 \Delta _{j} (t) + U _{x} ^{ij} (t) \Delta _{i} (t)} {a _{x} ^{2}} + \nonumber \\ +\frac{ U_{y} ^{mj} (t) \Delta _{m} (t) - 2 \Delta _{j} (t) + U _{y} ^{gj} (t) \Delta _{g} (t)}{a _{y} ^{2}} - \nonumber \\ - (1-T)(|\Delta _{j} (t)| ^{2} - \nu _{j})\Delta _{j} (t)\Bigg] + \tilde{f} _{j} (t), \qquad
\end{eqnarray}
\begin{eqnarray}
\mathcal{F} _{U _{x}} ^{jk} (t) \equiv -i(1-T)\mbox{Im}\{\Delta _{k} ^{*} (t) U _{x} ^{jk} (t) \Delta _{j} (t)\} - \nonumber \\ - \Big(\frac{\kappa ^{2}}{a _{y} ^{2}}\Big)[U _{y} ^{kn} (t) U _{x} ^{nm} (t) U _{y} ^{mj} (t) U _{x} ^{jk} (t) \cdot \nonumber \\ \cdot U _{y} ^{kh} (t) U _{x} ^{hg} (t) U _{y} ^{gj} (t) U _{x} ^{jk} (t) -1],
\end{eqnarray}
\begin{eqnarray}
\mathcal{F} _{U _{y}} ^{jm} (t) \equiv -i(1-T)\mbox{Im}\{\Delta _{m} ^{*} (t) U _{y} ^{jm} (t) \Delta _{j} (t)\} - \nonumber \\ - \Big(\frac{\kappa ^{2}}{a _{x} ^{2}}\Big)[U _{x} ^{ml} (t) U _{y} ^{li} (t) U _{x} ^{ij} (t) U _{y} ^{jm} (t) \cdot \nonumber \\ \cdot U _{x} ^{mn} (t) U _{y} ^{nk} (t) U _{x} ^{kj} (t) U _{y} ^{jm} (t) -1]. 
\end{eqnarray}

The relation of these variables to the magnetic induction is given by:
\begin{equation}
\mbox{B} _{z} = \frac{1}{i a_{x} a_{y}} (U _{x} ^{jk} U _{y} ^{kn} U _{x} ^{nm} U _{y} ^{mj} - 1).
\label{mean_field}
\end{equation}

The applied magnetic field and current are introduced in the sample through boundary conditions:
\begin{equation}
(\mbox{\bf{H}} _{2} - \mbox{\bf{H}} _{1})\times \hat{\mbox{\bf{n}}} = \mbox{\bf{K}},
\label{boundaryfield}
\end{equation}
Assuming vacuum outside, $\mbox{\bf{B}} _{2} = \mbox{\bf{H}} _{2}$. Therefore, the external influence of magnetic field and surface current are introduced by substituting the relevant product of link variables in the plaquette just outside the sample with the applied field and surface current.

Another boundary condition is necessary, the requirement of zero normal supercurrent at the border, leading to: 
\begin{equation}
\Delta _{j} = U _{x} ^{kj} \Delta _{k}.
\label{boundary_left}
\end{equation}

Once more, for proper replication and details concerning the discretisation of space and deduction of these results, one should consult \cite{lascio}. 

In order to obtain the time evolution of the magnetisation or a critical current curve as the applied magnetic field is varied, it is necessary to wait some time for the new value of the field, applied as boundary conditions, to reach the whole bulk of the sample, which evidently takes longer for bigger samples. This is a numerical restriction on the solution of the equations.

A physical restriction comes from the magnetic after-effect (see, for example \cite{getzlaff}), given cases in which the magnetisation reaction is delayed with respect to variations in the applied field. After an abrupt change in the field, such as those simulated by applying field steps, the magnetisation suffers an abrupt change as well, but continues to change slowly afterwards.

Both of these restrictions on the time necessary before moving on to the next step in applied field occur simultaneously and are inseparable in the framework of the numerical simulations. So, in order to achieve a stabilisation within a given step, one must wait the time necessary for evanescence of both numerical conformation and magnetic after effects. 

The state thus reached might resemble a plateau, with a constant level of magnetisation. Or it could be something more like an oscilatory steady state. In the case of type II superconductors, for given applied field and current, there is a vortex dynamics within the sample. In some cases, it might be observable as an oscilatory behaviour of the magnetisation, due to vortices moving in, out or within the sample, subjected also to surface barrier effects \cite{beanlivingston}. 

Therefore a proper algorithm should not be restricted to the reaching of constant values of the magnetisation, but instead it should be aimed at finding stationary states. So, it is henceforth defined the stabilisation of the step as the reaching of a stationary state (see, for example \cite{hamilton}) of magnetisation for given applied magnetic field and electric current.

Given that definition, unit root processes (such as a random walk) will be disconsidered for the purposes of the algorithm and the stationarity test will be restricted to testing the presence of a trend in data. Should any relevant unit root process appear, the algorithm is easily adapted including, for example, the augmented Dickey-Fuller test (\cite{dickey}, see also \cite{hamilton}). 

But most likely it is that the simulation is falling outside the limits of validity of either the solution methodology (in this case the link variable technique with simple Euler method) or the parameters of the TDGL equations. Should this not be the case, it could result possibly interesting new physics. But in any case, for finite time series samples, unit root and trend stationary tests can have arbitrarily low power \cite{cochrane}, meaning that trend stationary or unit root process can become indistinguishable on the basis of a finite sample. Therefore, the choice of trend stationary is based both on this empirical character and on a theoretical expectation of the sufficiency of local deterministic trends in the description of the evolution of the magnetic response of the material, until the reaching of stabilisation.

\begin{figure}[!htb]
\centering
\includegraphics[width=0.45\textwidth]{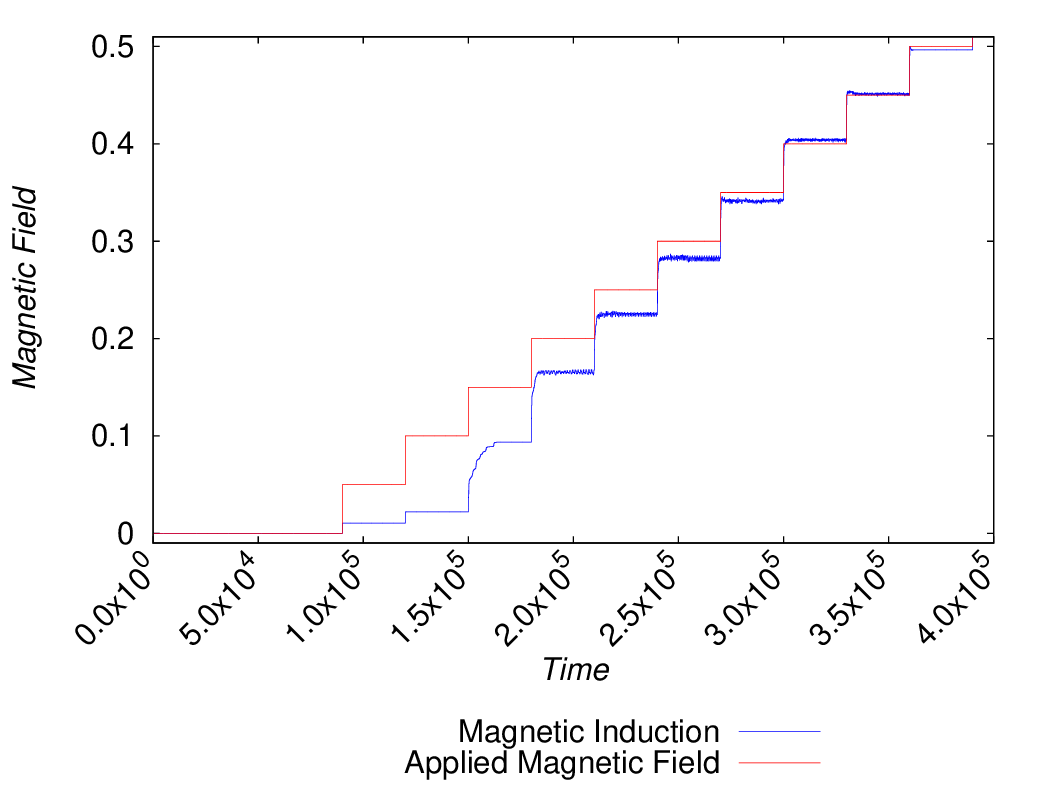}
\caption{(Colour online) Magnetic induction evolution with step = $2 \cdot 10 ^{6}$ iterations.}
\label{B_2106_1}
\end{figure}

The evolution in time of the magnetisation, or equivalently of the magnetic induction of the sample, is better understood following an example. In figure \ref{B_2106_1}, the applied magnetic field in time is shown, along with the averaged magnetic induction in the sample (whence one can calculate the magnetisation). 

The samples were taken as consisting of 95 points in each direction, with a total of 9025 points. The Ginzburg-Landau parameter was $\kappa = 2.0$, the lattice parameters were taken as $a _{x} = a _{y} = 0.5$, the normalised temperature $T = 0.5$, magnetic field step $\delta H = 0.05$, Euler algorithm interval $\delta t = 0.015$.

Each pair of values of applied magnetic field and current has been sustained for $2\cdot10^{6}$ iterations (the original algorithm proposed the use of $10^{5}$ as sufficient), this number will be refered to as the size of the step. In the first $4\cdot10^{6}$ iterations the applied magnetic field was held equal to zero, and the applied current was raised from zero to its final value (0.5) in two steps.

From the figure \ref{B_2106_1} one can immediately identify one important characteristic of the algorithm: the size of the steps should be variable. This is not the same as what is usually called adaptative time step, as the time step in the solution of the differential equations (Euler) is kept the same. It could also change, but that is not what is important here because it would give no information as to when could a value on the applied external influence be changed, since the response of the sample is already representative. 

To gain further insight, one may loosely classify the time evolution in four types. The first, observed in the interval $9\cdot10^{4}$ to $1.2\cdot10^{5}$ (in units of $t _{0}$) of figure \ref{B_2106_1}, shows a steep rise in $B$ and then a smooth steady state, with a time trend very close to zero. The second, from time $2.7\cdot10^{5}$ to $3.0\cdot10^{5}$, shows a similar situation, but in this case, before the steady state is reached there is a decrease in $B$ after the first steep rise (the $B$ values overshoot the equilibrium value). 

The third, time $1.8\cdot10^{5}$ to $2.1\cdot10^{5}$, is related to an oscilatory steady state. After the steep rise from the former step, the step is considered stabilised when it evolves around an average value with a definite standard deviation, in which case it follows a stationary time series. The fourth, observed in time interval from $1.5\cdot10^{5}$ to $1.8\cdot10^{5}$, evolves in a non stationary way for a great part of the step, showing a slower but persistent rise, which eventually, with enough iterations, reaches a plateau. In this step it is necessary to wait much more iterations than the first $10 ^{5}$ for the system to stabilise ($> 10 ^{6}$ iterations, the $3\cdot10^{4}$ units of time of the step are equivalent to $2\cdot10^{6}$ iterations).

Therefore one can see that a definition of stabilisation based on stationarity of time series is more aligned with the time evolution of the system, in a sense that the states in the final stage of evolution of magnetisation for a given value of applied field can all be accomodated within the definition. 

The output of an algorithm for identifying stabilisation in the framework of simulation of magnetisation is used to determine when to stop a given step and move on to the next, which amounts to let the simulation evolve for a number of iterations and iteratively test the stationarity hypothesis until it is reached. 

To assess the importance of accuracy in this case, it is relevant to evaluate the two possible outcomes of the choice of a wrong number of iterations.  

The first is that of a sub-optimal number of iterations. As can be observe from figure \ref{B_2106_1}, for many steps, even with $10 ^{4}$ iterations the stabilisation is already reached (for some it takes of the order of $10^{3}$ iterations). So the original prescription of $10 ^{5}$ iterations for each step wastes a great number of iterations, consuming a lot more time to finish simulations. 

\begin{figure}[!htb]
\centering
\includegraphics[width=0.45\textwidth]{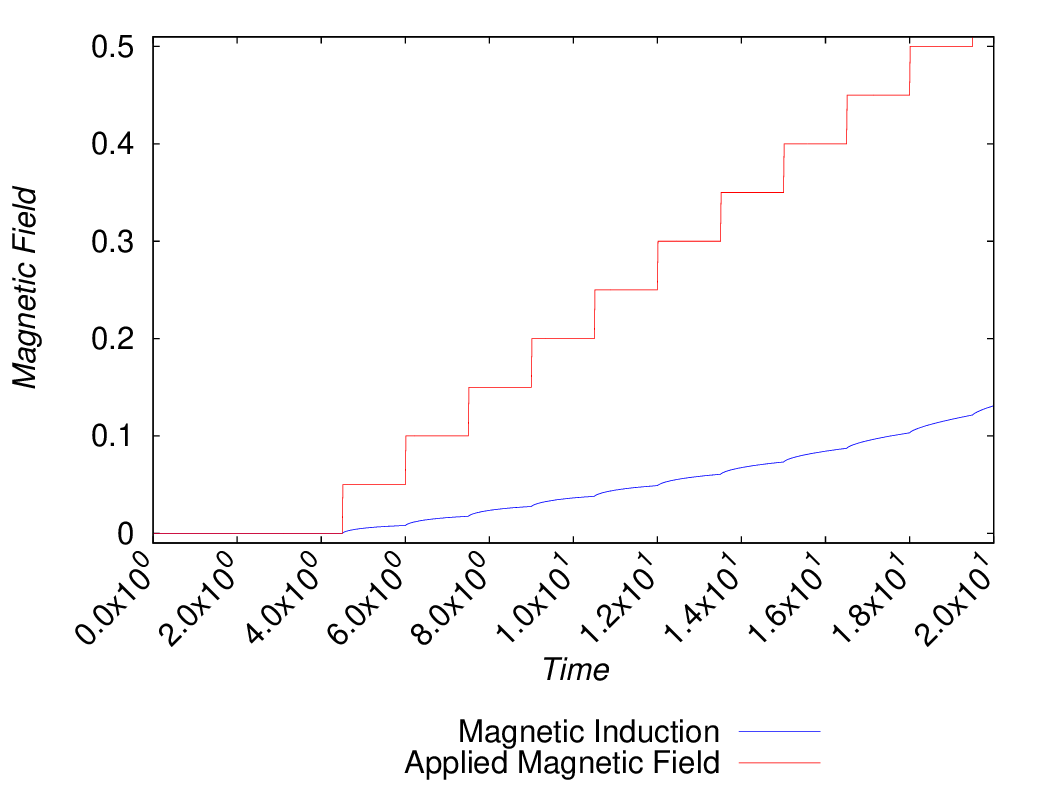}
\caption{(Colour online) Magnetic induction evolution with step = $10 ^{2}$ iterations.}
\label{B_102}
\end{figure}

The second consequence is far more dangerous, as not waiting enough iterations can lead to biases in the evolution of the field. The problem arises as the last value, or even an average, is taken as representative of the step, for a given applied field and current. To illustrate with an example, figure \ref{B_102} shows the extreme case of reducing the number of iterations to $10 ^{2}$. As can be seen in comparison with the evolution of figure \ref{B_2106_1}, the magnetic induction is severely biased if one takes the last value as representative of the step.

To compare distinct possible values of the magnetic response of the sample, independently of the number of iterations, it is more convenient to work with the magnetisation:

\begin{equation}
M = \frac{B_{avg} - H_{0}}{4\pi}.
\label{magnetisation}
\end{equation}

\begin{figure}[!htb]
\centering
\includegraphics[width=0.45\textwidth]{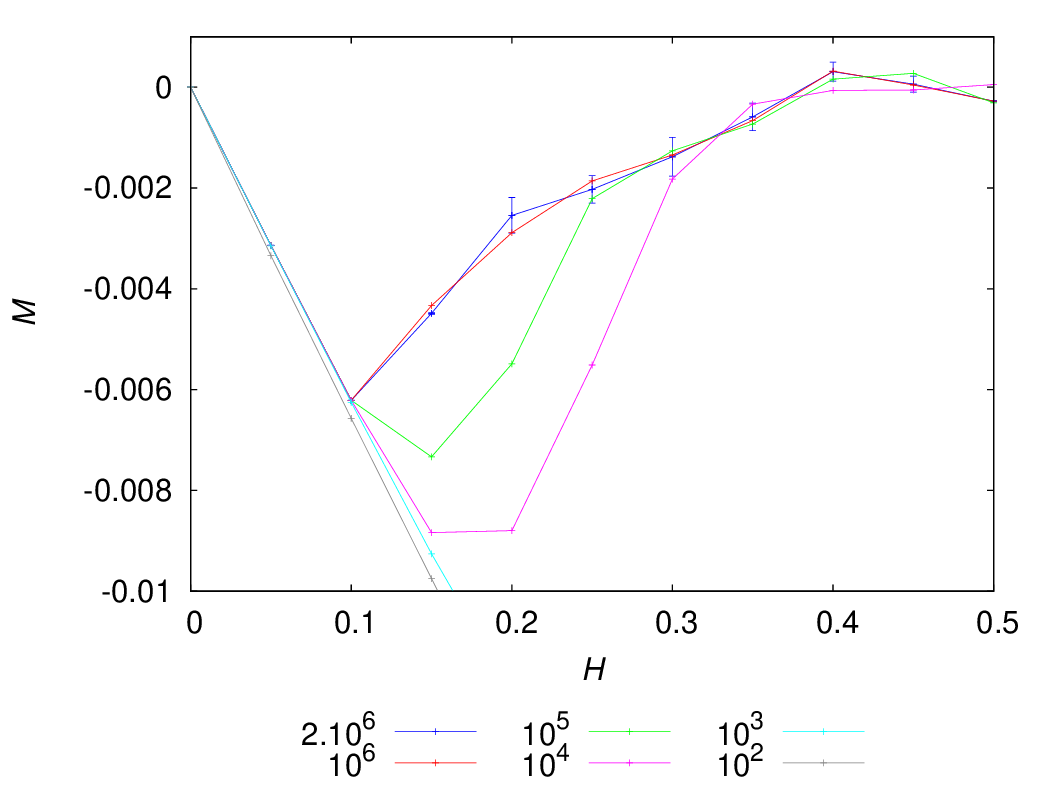}
\caption{(Colour online) Magnetisation for distinct step sizes.}
\label{M_ALL}
\end{figure}

In figure \ref{M_ALL}, the magnetisation is shown for the same sample simulated with step size from $10 ^{2}$ to $2\cdot 10 ^{6}$. Error bars are displayed only for the second half of the iterations for the case of step size $2 \cdot 10 ^{6}$. This is so because those values are the reference and the goal of the comparison is to evaluate if, substituting the reference value with those resulting of the alternative step size, the resulting values are compatible. Therefore, the variance of the other candidate simulations, with distinct step sizes, are of no relevance for this analysis. These error bars represent the confidence level of $0.999$, thus a z-score of 3.29053.

As can be seen, up to applied field of $H = 0.1$, step sizes of the order of $10 ^{3}$ are enough to avoid bias. From there up to about $0.25$, it takes roughly $10 ^{6}$ iterations. Therefore, the original proposed fixed step size of $10 ^{5}$ leads to significant bias in this interval. For $H \approx 0.3$, steps of the order of $10 ^{4}$ are sufficient. Hence, both at the beginning and at the end of the time evolution of the applied field, there is a waste of iterations spent on achieving stabilisation. 

Taking the above analysis and the generality of the concept of stationarity of time series as the criterium for ending a step in applied field or current, one can proceed to the formulation of the algorithm.

From a minimum number of iterations one can calculate a moving average and the corresponding variance. Also, identifying a local trend in data can give information related to the stage of evolution at each step. The variable step size until stabilisation is also addresed by the process of hypothesis testing of a local trend, as the step is ended as new values are reached that fall into the confidence interval of the given moving average. 

The algorithm is quite simple. In broad lines it is: at each new step in magnetic field or current, acumulate enough points in the time series, calculate the moving average of the average magnetic induction (or the magnetisation) until there is no significant trend in data, then move on to the next step.

\begin{figure}[!htb]
\centering
\includegraphics[width=0.45\textwidth]{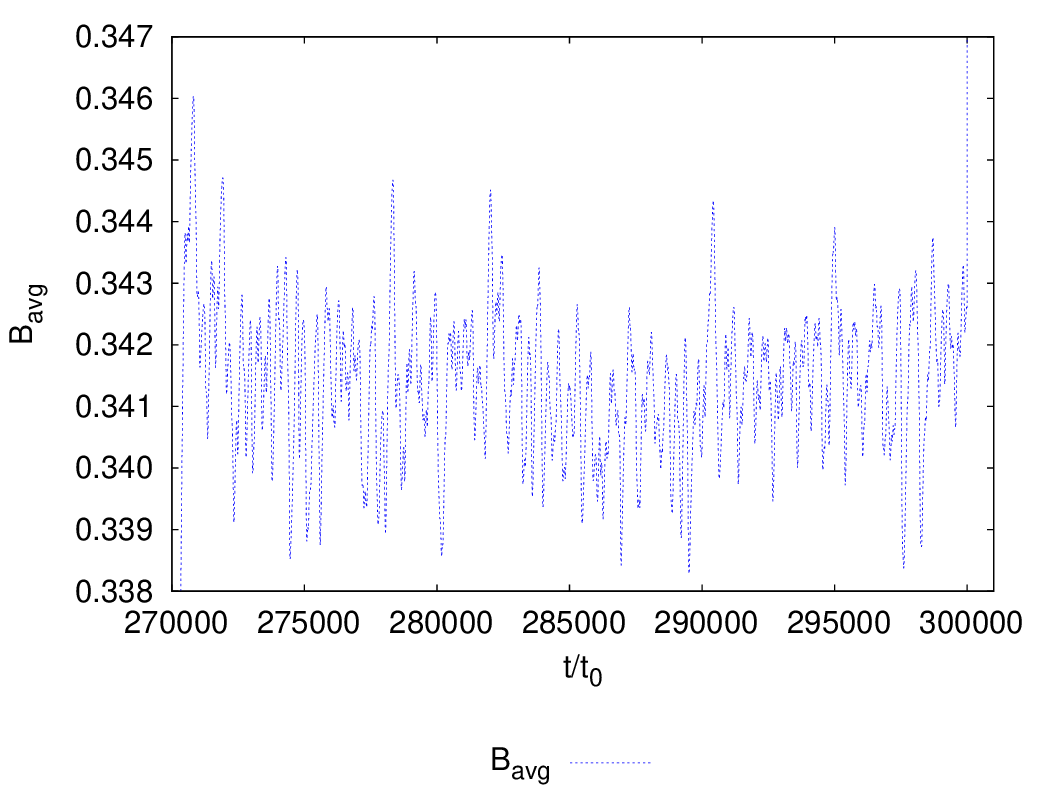}
\caption{(Colour online) Detailed view of the step $H = 0.35$ of the field of figure \ref{B_2106_1}.}
\label{peaks}
\end{figure}

That all is constrained to satisfy the two conditions: avoid bias and minimise the number of iterations necessary to stabilisation. In that, the details become important. The figure \ref{peaks} shows the details from figure \ref{B_2106_1}, for $H = 0.35$, whence one can observe that even with a moving average including $10 ^{3}$ points, due to the critical points in the successive peaks and to the multiple frequencies present in the oscillatory behaviour, a zero slope line can be found that is far from stabilisation, as the averages in successive samples of thousands of points can be of distinct levels, and smaller samples can easily lead to statistically significant differences in averages for adjacent time intervals.

To overcome this, for each step, there is a first trend calculated and when this reaches either a zero slope or a predetermined limit of iterations, another accumulation starts, to allow the calculation of a second moving average, to be considered representative of the steady state. Finding the first zero slope trend line helps to minimise the number of iterations, because if the first peak has been found faster, the algorithm will use less iterations to calculate the second moving average. If no peak has been found before the limit is reached, the processes within the sample are more complex and demand more iterations to be properly simulated (and effectively more time to reach stabilisation).

The significance level is considered the same in calculating both slopes and testing $H _{0}: \beta = 0$, where $B _{avg} = \alpha + \beta t$. As empirically observed, the algorithm is, in effect, little sensitive to variations in this value, but it should be high in order to be more conservative in finding the zero trend line. This is so because the null hypothesis is of a null coefficient, so the smaller the significance level, the harder it is to reject $H _{0}$, in which case we accept more easily that $\beta = 0$ and so the false negative rate tends to increase. 

The significance level is studied further on, but first it is necessary to introduce the algorithm in full details, showing all of its parameters and their significance, allowing for replication and eventually further discussions with other groups or even new improvements. 

Given a initial state (usually taken as a zero field cooled state) with zero applied field and current, apply the first step of field or current, depending on the simulation procedure desired, and start iterating the solutions to the TDGL equations following equations \ref{delta_it_eq} to \ref{uy_it_eq}, accumulating every point up to a lower limit $L _{1}$. 

It is instructive to try different possible values. Along the development of this study, the value of $L _{1} = 10 ^{4}$ has been found to be enough for the first moving average of the step. This value was obtained by calibration using many simulations, but it could be subjected to more calibrations still, should the need arise. Therefore, the justification for this particular value is taken {\it a posteriori} from the comparison with the resulting stabilisation values, but it can be considered relatively conservative.

The goal with this first sample of size $L _{1}$ is, besides increasing efficiency by removing from the analysis an initial steep increase, the minimisation of type II error, by avoiding the initial peak to which the curves are typically subjected after the steep increse that follows a change in the step. This mitigates the risk of not rejecting the null hypothesis in a part of the sample which tends to decrease and then to enter an oscillatory state. Depending on its size, in this region the moving window passes through points which give a zero slope line, simply because of the critical point present in the peak. In passing the first peak and reaching a typically oscillatory regime, the algorithm also becomes more efficient by cutting the initial phase of growth, which is evidently far from stabilisation. 

After this limit has been reached, start calculating the trend in time, that is, fit a function of the form:
\begin{equation}
B _{avg} = \alpha + \beta t
\label{trend}
\end{equation}
to the data. This should be repeated as every new point is obtained, for the time sample of fixed size $L _{1}$. As each new $\beta$ is calculated, it should be tested for significance. When deemed insignificant, or in the case of reaching an upper limit of total number of iterations for the first trend test in that step (not the value used in the moving average) $L _{2}$ with all coefficients calculated leading to rejection of $H _{0}$, the accumulation of points should start again, for the second moving average of the step. It could be taken $L _{2} = 10 ^{5}$, the value used in the present analysis, but it is also a little conservative, meaning spending possibly a little more iterations than strictly necessary to achieve stabilisation (but mitigating the risk of bias in a higher false negative rate).

This accumulation should be carried on until the time sample size (number of iterations available to calculate the trend) after the first part ended (the first iterations were discarded, for they probably carry a strong initial trend), is equal either to the number of the first iteration for which the trend was first compatible with zero or to the limit $L _{2}$. That will be the sample size for calculation of the trend test that is representative of the stabilisation.

Once this accumulation reaches the defined size, the window moves including every new point generated and excluding the most ancient of the time sample, again testing the significance of the time trend. The step in field or current ends when this trend is first found insignificant, or a second upper limit is reached (based on empirical studies, $L _{3} = 2 \cdot 10 ^{6}$ was used, but this is conservative and could be reduced depeding on the simulation). 

As the step is ended, the final value is taken to be representative of the average magnetic induction in the sample. This is better than to use the moving average as a representative value, because this avoids inconsistences between the magnetic induction calculated and the underlying order parameter and link variable values for every point in the discrete space. 

Then the next value of applied field or current in the simulation process is applied and the algorithm starts over again.

\section{Results}
\label{sec3}

\begin{figure}[!htb]
\centering
\includegraphics[width=0.45\textwidth]{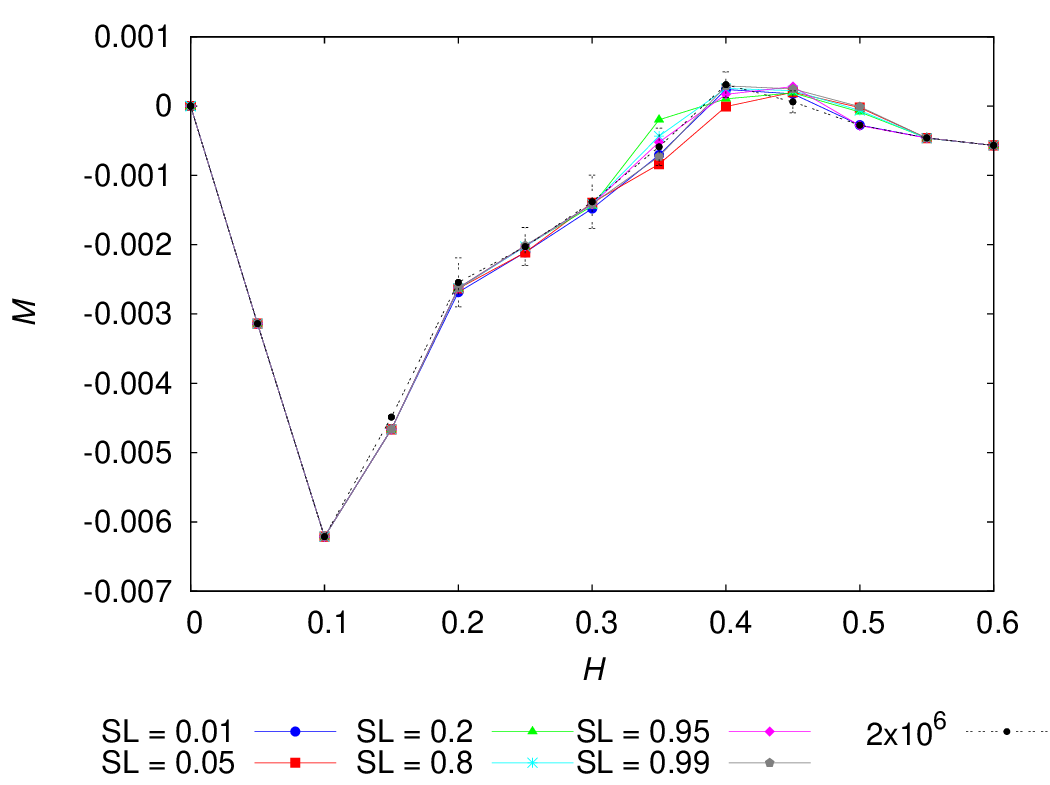}
\caption{(Colour online) Simulation with the proposed algorithm using distinct significance levels.}
\label{significance}
\end{figure}

Once the detailed algorithm was explained, allowing the replication of the results, it is important to evaluate the influence of the significance level chosen for hypothesis testing. Figure \ref{significance} shows the simulation of six different curves, for the same simulation parameters, with distinct significance levels, along with the reference values ($2\cdot10^{6}$ as step size, the error bars are the same as described for figure \ref{M_ALL}). The test applied was a student's t test, following the algorithm described above.

As can be observed, the difference among values from each curve, for the same value of applied field, is of little significance. One could take the significance level to 0.95 or higher (rejecting $H _{0}$ only if the p-value of the t test is greater than 0.95). However, due to the little difference observed, the simulations presented further on were calculated on the level of 0.8.

With the algorithm described and its parameters set, one can proceed to study the consequences of its application on the simulation process of superconducting samples, that is, validate the theoretical statistical deduction with simulated data.

As seen before, in figure \ref{B_2106_1}, with $2 \cdot 10 ^{6}$ iterations, steady state has been reached in all steps studied (up to and a little above the critical field). Further on, the corresponding values of magnetisation are considered as reference (the last value in the time series before moving to the next step).

The values of average magnetic induction obtained with the algorithm should be compared with the reference values. The intervals until stabilisation are much shorter however, difficulting the direct comparisons of the time series for $B _{avg}$. Therefore, once again it is best to recourse to comparisons between magnetisation values. 

\begin{figure}[!htb]
\centering
\includegraphics[width=0.45\textwidth]{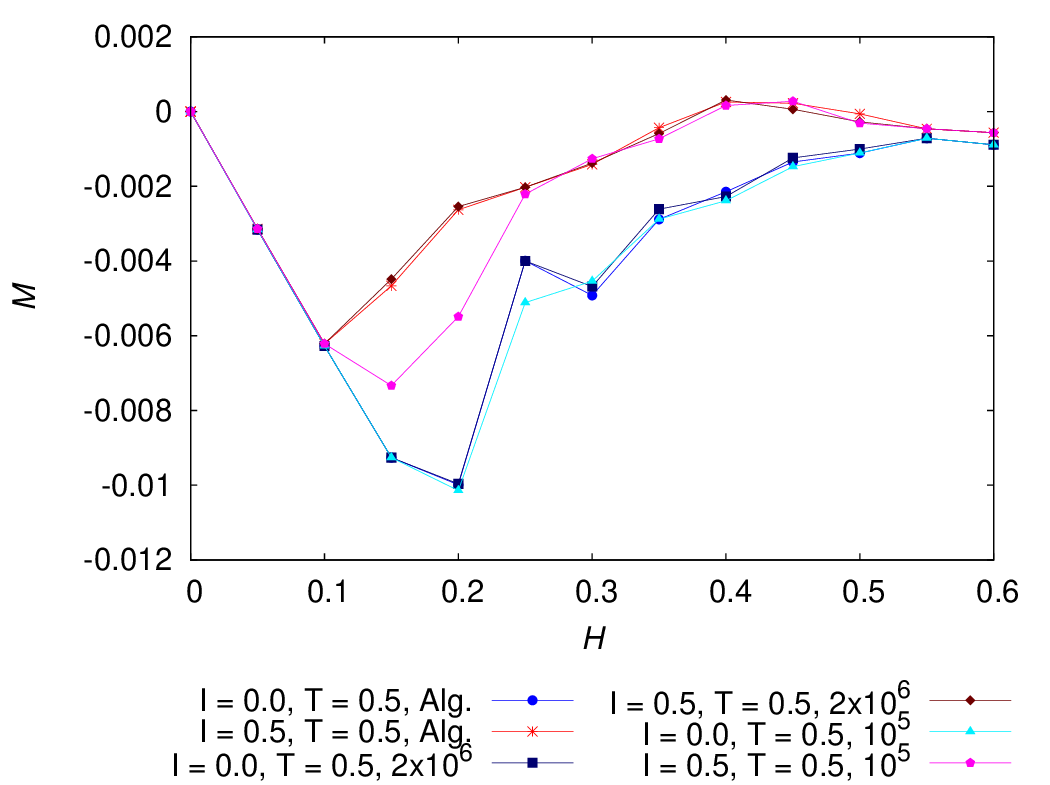}
\caption{(Colour online) Comparison of the proposed algorithm results, reference values and original step size, with and without applied current.}
\label{temperature_mag}
\end{figure}

Figure \ref{temperature_mag} shows the magnetisation curves obtained from the reference values, those obtained using the algorithm and the original prescription of $10 ^{5}$ iterations until stabilisation. These curves are presented for zero and 0.5 applied current. The temperature is maintained at $0.5$. Higher temperatures (up to 0.99) were evaluated throughout the study, but as these make the critical field and current much higher, the distinguishing features of a superconductor, regarding the magnetisation, tend to be much less pronounced, so the results were omitted. 

For the results of the proposed algorithm, all values of the magnetisation are very close to the reference. As for the original proposition of using $10^{5}$ iterations per step, it can be observed that it lead to bias. The presence of applied current tends to increase the level of bias, but this may be dependent of the characteristics of the sample, such as the presence of defects. Hence, one may infer that in studying critical currents it is crucial to avoid using a lower number of iterations (such as $10 ^{5}$) until stabilisation for some values of the applied magnetic field.

As for the other characteristic introduced by the new algorithm, the reduction in number of iterations and in time of execution, the table \ref{execution_times} shows the results for the comparison of the reference values ($2 \cdot 10 ^{6}$ iterations) with those from the new algorithm. Along with the number of iterations, it is shown the relative difference from the reference values (denoted {\it Diff. (ref.)} in table).

\begin{table}[h]
\caption{Detailed step sizes and relative differences from reference values for simulations with I = 0.5 at T = 0.5 with the proposed algorithm.}
\vspace{0.2cm}
\centering 
\begin{tabular}{cccc}
{\bf {\it H}} & {\bf {\it $B _{avg}$}} & {\bf {\it $N_{iter}$}} & {\bf {\it Diff. (ref.)}} \\ \hline
0.05 & 0.01 & 45131 & $< 0.01\%$ \\ \hline
0.10 & 0.02 & 55910 & $< 0.01\%$ \\ \hline
0.15 & 0.09 & 2000000 & $-2.38\%$ \\ \hline
0.20 & 0.17 & 343132 & $-0.64\%$ \\ \hline
0.25 & 0.22 & 227738 & $< 0.01\%$ \\ \hline
0.30 & 0.28 & 144285 & $-0.15\%$ \\ \hline
0.35 & 0.34 & 96010 & $0.57\%$ \\ \hline
0.40 & 0.40 & 159720 & $-0.17\%$ \\ \hline
0.45 & 0.45 & 73988 & $0.45\%$ \\ \hline
0.50 & 0.50 & 55087 & $0.54\%$ \\ \hline

\end{tabular}
\label{execution_times} 
\end{table}

As can be seen, the new algorithm reduces the bias to insignificant values from the reference. The total number of iterations for the steps presented in table \ref{execution_times} is 3329407, whereas with the original prescription of constant interval of $10 ^{5}$ iterations the total is 1200000, between a third and a half. However, the present algorithm has reduced bias considerably (the step for $H = 0.15$, with the original algorithm leads to an average magnetic induction of 0.058, that is little superior to half the value of reference 0.094). To obtain such reduction in bias with a constant interval, it would take $2.4 \cdot 10 ^{7}$ iterations, more than seven times the value obtained with the algorithm. One can see also that for many steps, the number of iterations necessary is reduced to below the original prescription.

Another important point is that with fine tunning of the parameters, including significance level and the size of the first moving window as well as the size of the second moving window, the algorithm could outperform even the original prescription, both in number of iterations and in bias reduction. However, the most serious point is the reduction of bias, which leads to more accurate simulations. So the values for the parameters that were presented in this article are those which give a robust prescription, leading to minimisation of bias without letting the number of iterations necessary for that becoming too high.

\section{Conclusion}
\label{sec4}

As was shown thoughout this article, the prescription of a constant interval for the stabilisation of the simulation step can lead to bias due to insufficient number of iterations and yet, in a same simulation, lead to a great number of iterations wasted unnecessarily.

The proposed algorithm was deduced and tested under extreme conditions considering the parameters of validity of the Time Dependent Ginzburg Landau equations. 

It solved the problem of bias introduced in the estimation of the magnetisation by exploring the fact that the intervals in which the time evolution of the average magnetic induction is non-stationary are limited and the characteristic of this non-stationarity is consistent with a deterministic time trend.

So, by introducing the concept of stationarity of the time series and statistical significance into the direct determination of the step size, it could be ensured that when the trend is insignificant the system has stabilised, and there is no need for spending further iterations on stabilisation, because the value reached should be statistically compatible with the value obtained from waiting a huge number of iterations. 

In other words, the algorithm is based on the fact that, when no time trend is significant in the evolution of magnetic induction, the stabilisation of the step has been reached.

This can be applied in obtaining more efficient and very importantly, unbiased, simulations, leading to an increased incentive for the use of the TDGL equations, which is desirable when substituting some approximate classical models of vortex dynamics, which in turn should lead to more precise conclusions, while at the same time, becoming more efficient.

\end{document}